\documentclass[journal=nalefd,manuscript=letter, layout=twocolumn]{achemso}
\usepackage{graphicx}
\usepackage{siunitx} 
\usepackage[usenames, dvipsnames]{color}

\author{S. Schaal}
\email{simon.schaal.15@ucl.ac.uk}
\affiliation{London Centre for Nanotechnology, University College London, London WC1H 0AH, United Kingdom}
\author{S. Barraud}
\affiliation{CEA, LETI, Minatec Campus, F-38054 Grenoble, France}
\author{J. J. L. Morton}
\affiliation{London Centre for Nanotechnology, University College London, London WC1H 0AH, United Kingdom}
\alsoaffiliation
{Department of Electronic \& Electrical Engineering, University College London, London WC1E 7JE, United Kingdom}
\author{M. F. Gonzalez-Zalba}
\email{mg507@cam.ac.uk}
\affiliation
{Hitachi Cambridge Laboratory, J.J. Thomson Avenue, Cambridge CB3 0HE, United Kingdom}

\title[CMOS classical-quantum circuit]{Conditional dispersive readout of a CMOS quantum dot via an integrated transistor circuit}

\abbreviations{CMOS,QD}

\begin{document}
%%%%%%%%%%%%%%%%%%%%%%%%%%%%%%%%%%%%%%%%%%%%%%%%%%%%%%%%%%%%%%%%%%%%%
%% The manuscript does not need to include \maketitle, which is
%% executed automatically.  The document should begin with an
%% abstract, if appropriate.  If one is given and should not be, the
%% contents will be gobbled.
%%%%%%%%%%%%%%%%%%%%%%%%%%%%%%%%%%%%%%%%%%%%%%%%%%%%%%%%%%%%%%%%%%%%%
\begin{abstract}
Quantum computers require interfaces with classical electronics for efficient qubit control, measurement and fast data processing. Fabricating the qubit and the classical control layer using the same technology is appealing because it will facilitate the integration process, improving feedback speeds and offer potential solutions to wiring and layout challenges. Integrating classical and quantum devices monolithically, using complementary metal-oxide-transistor (CMOS) processes, enables the processor to profit from the most mature industrial technology for the fabrication of large scale circuits. Here we demonstrate the integration of a single-electron charge storage CMOS quantum dot with a CMOS transistor for control of the readout via gate-based dispersive sensing using a lumped element $LC$ resonator. The control field-effect transistor (FET) and quantum dot are fabricated on the same chip using fully-depleted silicon-on-insulator technology. We obtain a charge sensitivity of $\delta q=165\, \si{\micro e  Hz^{-1/2}}$ when the quantum dot readout is enabled by the control FET. Additionally, we observe a single-electron retention time of the order of a second when storing a single-electron charge on the quantum dot at milli-Kelvin temperatures. These results demonstrate first steps towards time-based multiplexing of gate-based dispersive qubit readout in CMOS technology opening the path for the development of an all-silicon quantum-classical processor.
\end{abstract}

%\section{Introduction}

Multiple quantum computing platforms have already reached the level of few-qubit demonstrators~\cite{Barends2016, Schindler2013} and are addressing the challenges of scaling up to larger arrays in order to implement error-correction protocols~\cite{Kelly2015, Corcoles2015, Riste2015} and tackle practical problems. In each case, interfaces are required between classical control systems (which may include optics, microwaves and DC electronics, depending on the technology platform) to perform control and readout of the quantum state of the system~\cite{Reilly2015}, including low-level operations to implement feedback and error correction, and high-level operations to perform the quantum algorithm. 

%, it would be unfeasible to wire each qubit separately. Hence, operating these complex circuits will require interfacing the quantum machinery with classical electronics for accurate control and readout of the quantum state of the system~\cite{Reilly2015}.

Amongst the most promising candidates for large-scale quantum computing are electron spins in semiconductor quantum dots, particularly in isotopically purified silicon~\cite{Schreiber2014,Veldhorst2014,Veldhorst2015,Eng2015}. Silicon is attractive as a host material as it offers long coherence times and a variety of qubit implementations ($T_2^e=40\, \si{\micro s}$ in Si/SiGe~\cite{Kawakami2014}, $T_2^e=28\, $ms in MOS~\cite{Veldhorst2015} and $T_2^e=550\, $ms in donor~\cite{Muhonen2014} based nanostructures) and coupling geometries~\cite{Veldhorst2015,Eng2015,Urdampilleta2015,Zajac2016,Mi2017,Hutin2016}. These silicon-based approaches can all, to varying degrees, leverage nanofabrication techniques used in the semiconductor industry, and it is also possible to directly make use of CMOS technology (responsible for an exponential growth of transistor count in classical processors~\cite{Waldrop2016}) as the basic platform for qubit devices~\cite{Urdampilleta2015,Hutin2016}. The small footprint of the qubit nanostructures themselves would allow for high-density integration of the qubits, in principle~\cite{Zajac2016}, however, exploiting this potential to scale up to a large number of densely packed qubits brings formidable challenges in qubit addressing. 

%into practice has proven to be challenging due to the small inherent tolerance levels of quantum states~\cite{Schreiber2014}.

% Semiconductor quantum dots in silicon devices are promising candidates for quantum computation \cite{Veldhorst2014,Veldhorst2015,Urdampilleta2015,Hutin2016}. Additionally, single dopants in the same material are thought to realize excellent memories\cite{Morton2008a,Tyryshkin2011,Maurer2012,Steger2013,Saeedi2013,Wolfowicz2013}.
% The small footprint of such nanostructures allows a high density\cite{Zajac2016}. However, bringing the vision of scaling to a processor with a large number of qubits into practice has proven to be challenging due to the small inherent tolerance levels of quantum states\cite{Schreiber2014}.

%In classical electronics, the complementary metal-oxide-semiconductor (CMOS) platform gave rise to an exponential growth of transistor count~\cite{Waldrop2016}. For quantum technologies, CMOS processes could provide the necessary level of reproducibility combined with easy integration of control and readout electronics for large scale quantum processors~\cite{Clapera2015}. 
%Moreover, silicon CMOS integrated circuit have undergone decades of advances leading to an unparalleled level of integration. 

CMOS technologies provide a natural route towards tackling challenges in qubit addressing and the integration of control and readout electronics for large scale quantum processors~\cite{Clapera2015}. This is reflected in a recent proposal by Veldhorst \textit{et al.}, which considered on-chip integration of quantum and classical hardware, with a CMOS-based quantum processor relying on quantum-dot spin qubits and transistor-based control circuits together with charge storage and a scalable gate-based readout scheme~\cite{Veldhorst2016}. The architecture has similarities with the floating memory gates found in modern DRAM chips \cite{Keeth2008}. In both cases a key concept which underpins scalability is multiplexing: the ability to address arrays of $2^n$ (qu)bits using O($n$) leads.

%Recently, it has been suggested that these concepts can also be employed to create a scalable qubit system\cite{Veldhorst2016} relying on the CMOS platform. The architectural key idea is to integrated CMOS based quantum dot spin qubits on-chip with transistor control and readout circuits with floating memory gate electrodes in analogy to the DRAM architecture. 

On-chip multiplexing circuitry to address elements of an array of gate-defined quantum devices has been demonstrated in GaAs~\cite{Smith2015,Smith2015a} (256 QPCs) and Si/SiGe~\cite{Ward2013} (four quantum dot devices). Similarly, a switching matrix for a high-frequency transmission line has been realized~\cite{Hornibrook2015} showing routes towards controlling large scale devices.
In addition to control, fast high-fidelity readout is another essential requirement, and for quantum dot devices this is commonly achieved using nearby electrometers\cite{Elzerman2004b,Reilly2007a,Barthel2010}. Gate-based readout~\cite{Colless2013,Betz2015,Gonzalez-Zalba2016} provides a more scalable alternative, taking the gate(s) that define the quantum dot and using them additionally as a sensor. For both separate and gate-based qubit readout, sensitivity and speed is improved by using radio frequency (rf) techniques: coupling the sensor to a rf resonant circuit. Recently, gate-based approaches have reached a sensitivity of $37 \, \si{\micro e  Hz^{-1/2}}$~\cite{Gonzalez-Zalba2015}, comparable to rf electrometers~\cite{Aassime2001,Mason2010,Reilly2007a}. Frequency-domain multiplexing is a useful method to read out multiple sensors simultaneously, however, the scalability of this approach is limited by the accessible bandwidth~\cite{Hornibrook2013}.

Here, we report on gate-based rf-reflectometry of a quantum dot, controlled through a field-effect transistor (FET) at milli-Kelvin temperature.
Both the quantum dot device and control FET are fabricated using the same CMOS fabrication process and are realized on the same chip as envisioned by Veldhorst \textit{et al.}~\cite{Veldhorst2016}.
Our experiment realizes a first step towards an integrated time-based multiplexing of gate-based reflectometry readout (CMOS and cryogenic compatible) by demonstrating sensitive readout through a control FET in the `ON' state combined with floating gate charge storage in the `OFF' state. 
%We combine this with charge storage time-based multiplexing to allows subsequent readout of gate-based sensors using a single resonant circuit.
%
%We first characterize the quantum device through transport measurements, before moving to RF-based readout, controlled by a FET, including an analysis of the charge sensitivity. We finish by characterizing the device dynamics. 
%

%The concept of multiplexing is a key aspect for any type of scalable electronic platform and enables to address $n$ outputs with only $\sqrt{n}$ addressing lines.
%For memory devices these lines are usually refereed to as word and data (bit) line. 
%In order to allow subsequent readout of gate-based sensors in a scalable way multiplexing needs to be combined with gate charge locking. 

%This experiment is motivated by the DRAM architecture.%which is based on a capacitive data storage combined with a transistor based control circuit. 
%\citet{Veldhorst2016} showed that such an approach based on transistor control circuits and floating memory gate electrodes can also provide a scalable quantum computer architecture even though a qubit requires much greater tunability combined with much smaller tolerance levels compared to classical information processing.

%\section{Experimental setup and DC characterisation}

\begin{figure}[th!]
  \includegraphics[width=\linewidth]{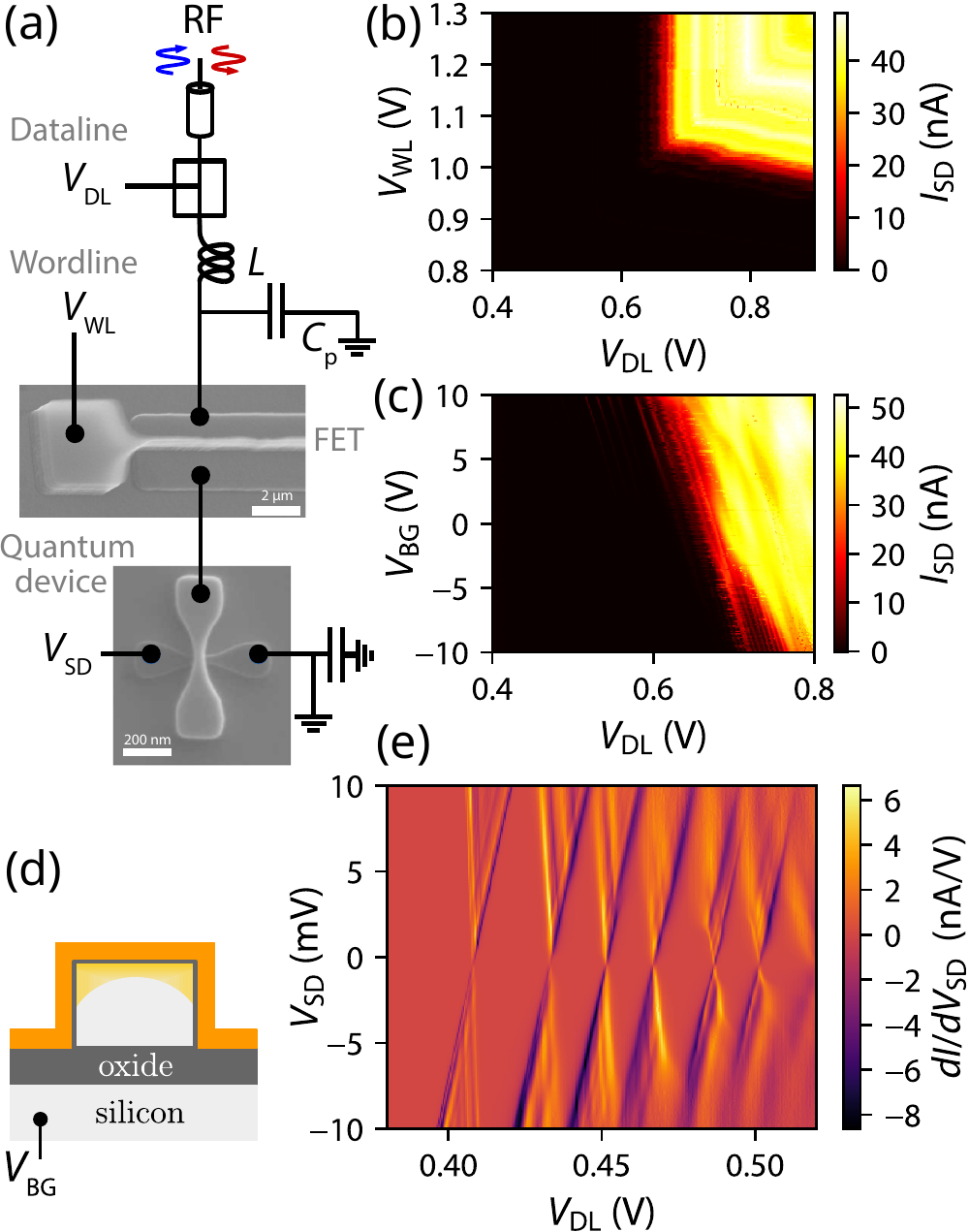}
  \caption{{\bf Experimental setup and DC transport measurements}. \textbf{(a)} Measurement circuit schematic, including SEM micrographs of the control FET and quantum device. Control and measurement signals are sent to the quantum device via the channel of a control FET.
  %The radio-frequency response is probed by connecting a resonant circuit to thesource of the FET. With a bias tee an additional voltage $V_{\mathrm{DL}}$ is applied. Both signals are selectively passed onto the qubit gate depending on the control FET gate voltage $V_{\mathrm{WL}}$. 
 \textbf{(b)} Transport through the quantum device as a function of $V_{\mathrm{DL}}$ and $V_{\mathrm{WL}}$ yielding the threshold voltage of the control FET and quantum device at $V_{\mathrm{BG}}=0\, $V. \textbf{(c)} Turn-on characteristic of the quantum device as a function of $V_{\mathrm{BG}}$ when the FET is biased well above threshold at $V_{\mathrm{WL}}=1.3\, $V. \textbf{(d)} Cross-section illustration of the nanowire-based quantum device under high back-gate bias and near-threshold top-gate bias, such that a single quantum dot forms. \textbf{(e)} Coulomb diamonds indicating a single quantum dot in the quantum device at $V_{\mathrm{BG}}=10\, $V.}
  \label{fig:setup&dc}
 \end{figure}
 
Both the quantum dot device and control FET are based on CMOS transistors fabricated on a silicon-on-insulator (SOI) substrate where the silicon substrate acts as a back-gate. The control FET is realized using a wide channel ($10\, \si{\micro m}$) and short gate ($50\, $nm) device, while the quantum device consists of a narrow nanowire ($60\, $nm) with a short gate ($30\, $nm). The measurement setup is depicted in Fig.\@ \ref{fig:setup&dc}.(a) including a SEM micrograph of both devices. The connection between the devices is made on-chip using a short bond wire.
In the transistor with the nanowire channel, we expect formation of quantum dots in the upper corners of the nanowire due to an enhanced field effect under the gate\cite{Voisin2014}. 
At large positive back-gate voltage the wave-function of electrons in the corners is expected to extend further into the center of the wire resulting in a single extended quantum dot (see  Fig.\@ \ref{fig:setup&dc}.(d))~\cite{Betz2014}.

In this configuration, the combined quantum-classical CMOS circuit has two primary inputs which, in analogy to a multiplexer or memory device, we refer to as the word- and data- (bit) line. The wordline is connected to the gate of the control FET, while the dataline passes through the channel of the control FET and is applied to the gate of the quantum device. Source-drain transport through the quantum device can be measured directly, or readout based on rf-reflectometry can be performed by applying rf-modulation onto the dataline (via an on-PCB bias tee) and using an $LC$ resonant circuit made from a surface mount inductor and the parasitic capacitance of the device $C_\mathrm{p}$. 
In this way, the rf-modulation and dataline voltage $V_{\mathrm{DL}}$ should only be applied to the quantum device gate when the control FET gate voltage $V_{\mathrm{WL}}$ is above threshold. The $LC$ resonator response is amplified at multiple stages, followed by IQ-demodulation (not shown, see Gonzalez-Zalba \textit{et al.}~\cite{Gonzalez-Zalba2016} for details) from which the amplitude and phase of the reflected signal is obtained. The phase $\Phi$ of the reflected signal is sensitive to small changes $\Delta C$ in the quantum capacitance of the quantum device, associated, for example, with the tunneling of single electrons: $\Delta \Phi\approx -\pi Q\Delta C/C_\mathrm{T}$ with $Q$ being the quality factor of the resonator and $C_\mathrm{T}$ being the total capacitance of the circuit.

First, we characterize the quantum device and control FET through transport measurements. We measure the source-drain current through the quantum device as a function of $V_{\mathrm{DL}}$ and $V_{\mathrm{WL}}$, under some small source-drain bias ($V_\text{\rm{SD}}=1\, $mV), observing the turn-on of the FET and quantum device in Fig.\@  \ref{fig:setup&dc}.(b).  
When the control FET is operated below threshold (the `OFF' state), the gate of the quantum device is isolated from the the signal on the dataline. In this state of the circuit, the quantum device gate floats, allowing it to retain its charge over a timescale of a second, as we explore later on. For measurements where $V_{\mathrm{WL}}$ is ramped slowly (as in Fig.\@  \ref{fig:setup&dc}.(b)), the quantum device gate voltage tends to $0\, $V when the control FET is `OFF'.
Once the control FET is operated well above threshold the transfer curve of the quantum device transistor can be measured, while a transition region is also apparent where the control FET is still strongly resistive.
From Fig.\@  \ref{fig:setup&dc}.(b) we estimate the threshold voltage of the quantum device $V^\mathrm{Q}_{\mathrm{th}}=0.63\, $V and the FET $V^\mathrm{FET}_\mathrm{th}=0.37\, $V (at $V_\mathrm{BG}=0\, $V).
The control FET threshold voltage is calculated as  $V^\mathrm{FET}_\mathrm{th}=V_\mathrm{WL} - V_\mathrm{DL}$ at $(V_\mathrm{WL}, V_\mathrm{DL})=(1.02, 0.65)\, $V and additionally depends on $V_\mathrm{BG}$ (not shown).

An important tuning parameter for the quantum device used here is the back-gate voltage $V_{\mathrm{BG}}$ applied to the substrate --- in principle, this affects both the control FET and quantum device as they are realized on the same chip. In Fig.\@ \ref{fig:setup&dc}.(c), we demonstrate that the control FET remains in a low-resistance state for a large range of $V_{\mathrm{BG}}$ when operated well above threshold ($V_{\mathrm{WL}}=1.3\, $V). At large positive back-gate voltage we observe clear and regular Coulomb oscillations.
Finally, in Fig.\@ \ref{fig:setup&dc}.(e), we confirm the formation of a single few-electron quantum dot under the gate of the quantum device by measuring Coulomb diamonds at $V_{\mathrm{BG}}=10\, $V and $V_{\mathrm{WL}}=1.3\, $V. We observe a first addition energy of about $16\, $meV showing strong confinement compatible with previous measurements~\cite{Voisin2014,Urdampilleta2015}.

%\section{RF and charge sensitivity}

% Start of Figure 2
\begin{figure}[t]
  \includegraphics[width=\linewidth]{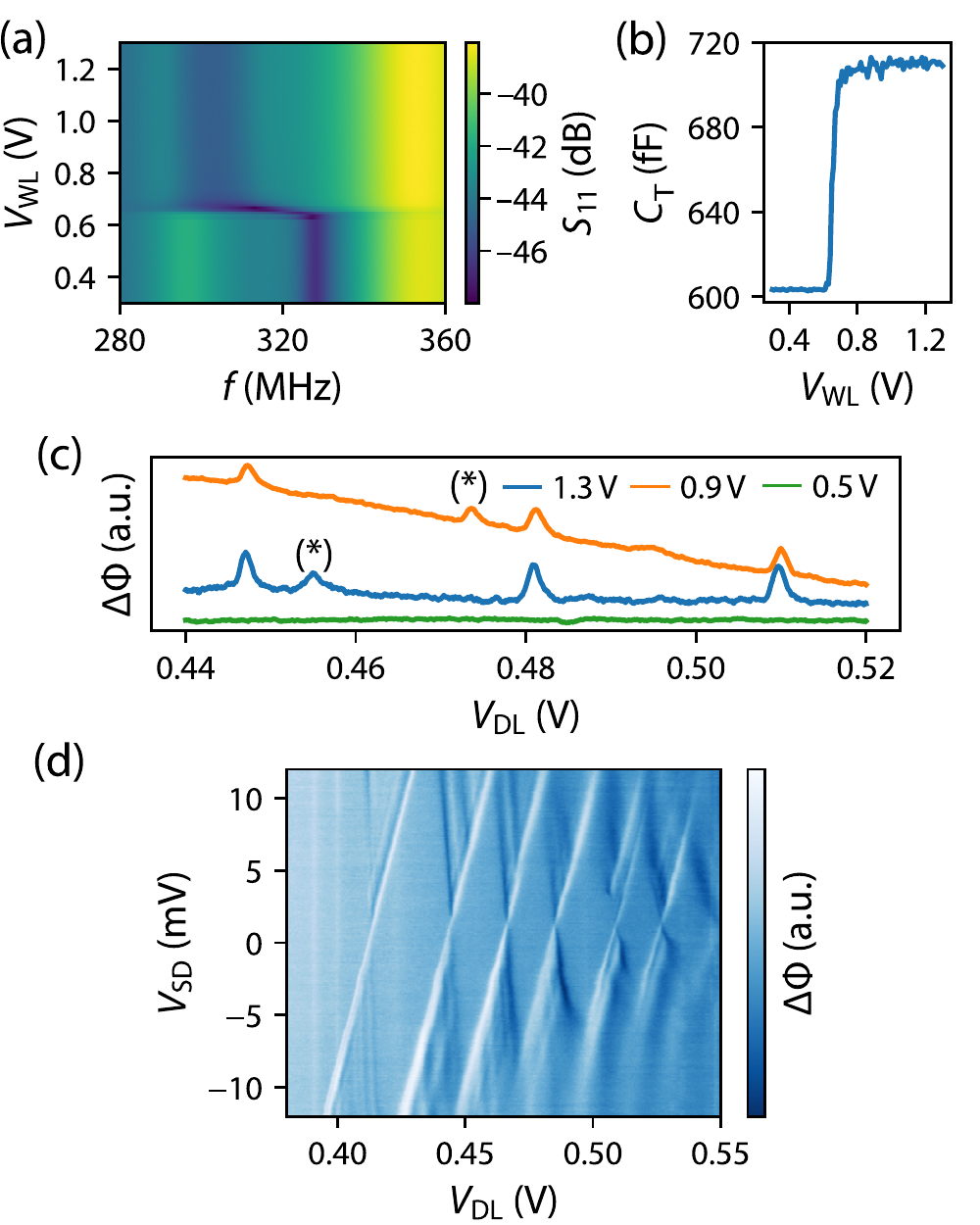}
  \caption{{\bf RF characterization and charge sensitivity}. \textbf{(a)} $S_{11}$ of the rf circuit as a function of $V_{\mathrm{WL}}$ (with $V_{\mathrm{DL}}=0.4\, $V and  $V_{\mathrm{BG}}=10\, $V). \textbf{(b)} Total resonator capacitance $C_\mathrm{T}$ as a function of $V_{\mathrm{WL}}$ with $L=390\, $nH.
  %A shift in the resonance frequency is seen when the control FET is operated above threshold. JM- straight repetition of text
%  \textbf{(b)} Parasitic capacitance as a function of $V_{\mathrm{WL}}$.
%
%JM COMMENT: {\color{blue} Not sure about panel (b) - firstly, $C_p$ is meant to be constant, with $C_{FET}$ being added to it, so the label is confusing. Secondly, where does this come from - if it is just calculated from the shift in frequency, then it's not really worth bothering with. We can just state the shift in capacitance. I would delete panel b and increase with width of panel a}
  \textbf{(c)} Change in phase response for different $V_{\mathrm{WL}}$
  showing three Coulomb oscillations only when the control FET is operated above threshold. Features originating from charge transitions within the control FET itself are indicated as ($\star$). \textbf{(d)} Coulomb diamonds measured in the phase response ($V_{\mathrm{BG}}=10\, $V and $V_{\mathrm{WL}}=1.3\, $V).}
  \label{fig:rf}
\end{figure}
% end of Figure 2

We now move on to performing gate-based rf readout of the quantum dot, and evaluating the achievable charge sensitivity, considering the potential impact of the additional parasitic capacitance and dissipation from the control FET circuit.
First, we characterize the $LC$ resonant circuit by measuring reflection ($S_{11}$) as a function of $V_{\mathrm{WL}}$ (see Fig.\@ \ref{fig:rf}.(a)). We observe lowering of the resonance frequency when the control FET is operated above threshold ($V_\mathrm{WL}> 0.63\, $V) due to the additional capacitance of the FET circuit that appears in parallel to $C_\text{p}$. From Fig.\@ \ref{fig:rf}.(b) which shows the total capacitance $C_\mathrm{T}$ of the $LC$ circuit (obtained from Fig.\@ \ref{fig:rf}.(a) using the nominal inductance $L=390\, $nH) we estimate the contribution of the FET circuit to $105\, $fF. 

Next, we examine the phase response of the resonant circuit as a function of the gate voltage on the control FET (see Fig.\@ \ref{fig:rf}.(c)), using rf modulation at frequency $f_{\mathrm{rf}}=313\, $MHz and power $P_{\mathrm{rf}}=-88\, $dBm. Starting with the control FET well above threshold ($V_{\mathrm{WL}}=1.3\, $V), in the strong accumulation regime, we observe three principal Coulomb peaks when ramping $V_{\mathrm{DL}}$ (blue trace). The peaks remain initially visible as $V_{\mathrm{WL}}$ is reduced, though a background signal begins to dominate as the control FET enters the weak inversion regime where the FET gate capacitance strongly depends on $V_\text{WL}-V_\text{DL}$. 
Since $V_\text{DL}$ is modulated by the rf-signal, this is picked up in the dispersive response of the resonator as an additional change in capacitance that in turns produced and additional phase shift that depends on $V_\text{DL}$. Eventually, when $V_\text{WL}<0.5\, $V the control FET is below threshold and the dispersive response vanishes (green trace). We note the appearance of additional features in the scan (indicated by asterisks) which we identify with single-electron tunneling events in the control FET due to their $V_{\mathrm{WL}}$ dependence. These features become more apparent when operating the control FET close to threshold. Fig.\@ \ref{fig:rf}.(d) shows rf measurements (with the control FET well above threshold) showing Coulomb diamonds of the quantum dot in the same voltage region as the transport measurements in Fig.\@  \ref{fig:setup&dc}.(e). The correspondence between both sets of measurements shows that, in the strong accumulation regime, the FET channel has negligible impact on the rf readout. 

\begin{figure}[t]
  \includegraphics[width=\linewidth]{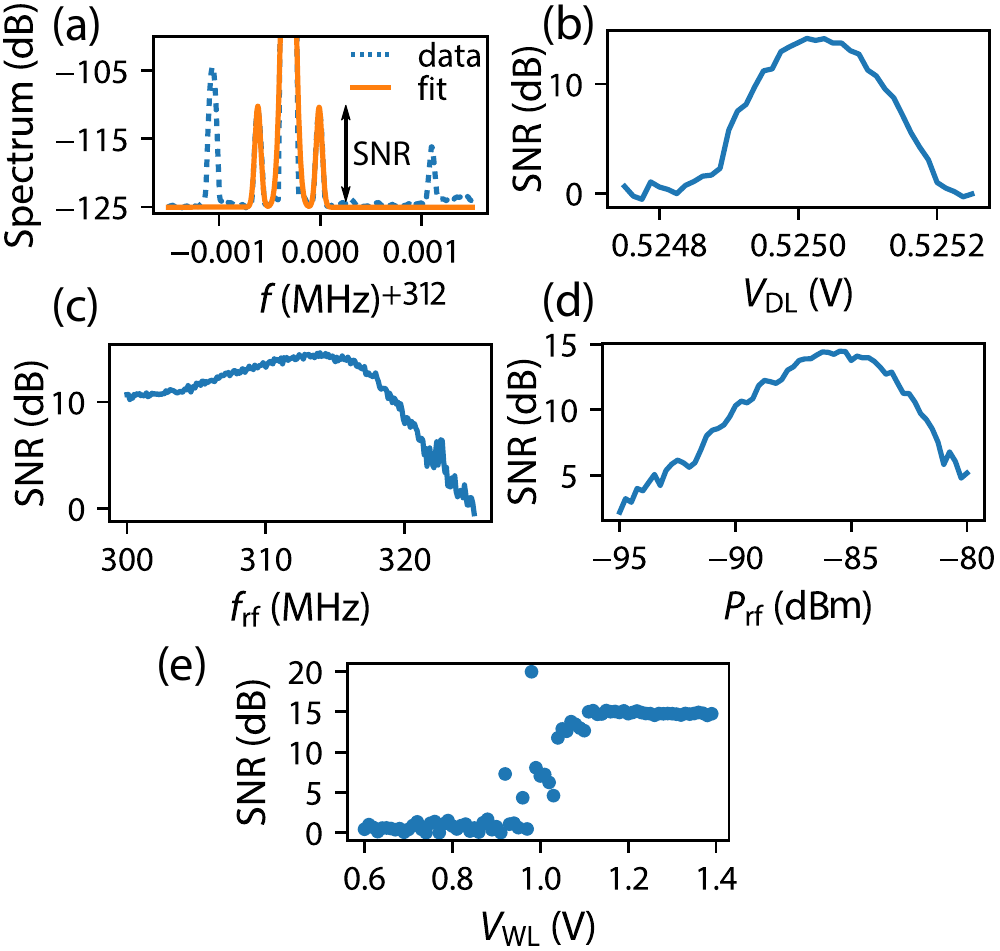}
  \caption{{\bf Charge sensitivity of the gated rf readout.} \textbf{(a)} Sidebands in the spectrum when operating at the point of maximum slope of a Coulomb oscillation with an equivalent excitation of $0.01\, $e at $303\, $Hz superimposed on the dataline. Signal-to-noise-ratio (SNR) as a function of \textbf{(b)} dataline DC voltage $V_{\mathrm{DL}}$, \textbf{(c)} carrier frequency $f_{\mathrm{rf}}$, \textbf{(d)} carrier power $P_{\mathrm{rf}}$, and \textbf{(e)} FET gate voltage $V_{\mathrm{WL}}$. When not being swept, the following parameter values are used:
$V_{\mathrm{WL}}=1.3$~V,
$V_{\mathrm{DL}}=0.525$~V, 
$f_{\mathrm{rf}}=312$~MHz, 
$P_{\mathrm{rf}}=-85$~dBm}
  \label{fig:snr}
\end{figure}

To measure the charge sensitivity of the gate-based sensor with control FET, we  apply a small-amplitude signal of frequency $f_\text{s}=303\, $Hz (in addition to the rf-modulation at $f_{\mathrm{rf}}$) onto the dataline and monitor the signal-to-noise ratio (SNR) in dB of the sidebands appearing in the frequency spectrum at $f_{\mathrm{rf}} \pm f_\text{s}$. The amplitude of the signal ($0.2\, $mVpp) corresponds to a change of $\Delta q=0.01e$ in the charge on the quantum dot, where $e$ is the charge of the electron.
A typical spectrum in shown in Fig.\@ \ref{fig:snr}.(a).
%for $f_\text{s}=303\, $Hz and a signal amplitude of $\Delta q=0.01e$, where $e$ is the charge of the electron. 
We optimize the sideband SNR by tuning the circuit parameters $V_\text{DL}$, $f_\text{rf}$ and $P_\text{rf}$ as seen in  Fig.\@ \ref{fig:snr}.(b-e) respectively. First, we find the maximum sensitivity at the point of maximum slope in the response of the resonator, $V_\text{DL}=0.525$~V. 
The rf-frequency dependence, in Fig.\@ \ref{fig:snr}.(c), reveals a maximum at $f_\text{rf}=313$~MHz and a 3dB bandwidth of 13~MHz which translates in to a loaded Q-factor of 24 in the `ON' state of the control FET. This contrasts with `OFF'-states measurements, and previous results~\cite{Gonzalez-Zalba2015} where the loaded $Q$ was $\approx 40$.
The optimal value for the rf power $P_\text{rf}$ was found to be $-86$~dBm. 
Finally, observing the SNR as a function of $V_\mathrm{WL}$ (Fig.\@ \ref{fig:snr}.(e)) we identify two plateaus corresponding to the `ON'- and `OFF'-states of the control FET. In the approximately linear transition between the plateaus, we observe multiple scattered data points which we attribute to transitions in the weak inversion regime of the FET (c.f.\ starred features in Fig.\@ \ref{fig:rf}.(c)). 
Overall, using optimized circuit parameters we obtain a SNR of $15.6\, $dB which translates into a charge sensitivity of $\Delta q/ (\sqrt{2 B_\mathrm{SA}}\times 10^{\mathrm{SNR}/20})=165\, \si{\micro e  Hz^{-1/2}}$ for the chosen spectrum analyzer bandwidth $B_\mathrm{SA}=50\, $Hz. This result compares well to rf-QPC devices\cite{Mason2010} and demonstrates a near two orders of magnitude improvement compared to GaAs based gate sensors\cite{Colless2013} and comes close to a previously reported sensitivity in a similar device\cite{Gonzalez-Zalba2015}.

\begin{figure}[th!]
  \includegraphics[width=\linewidth]{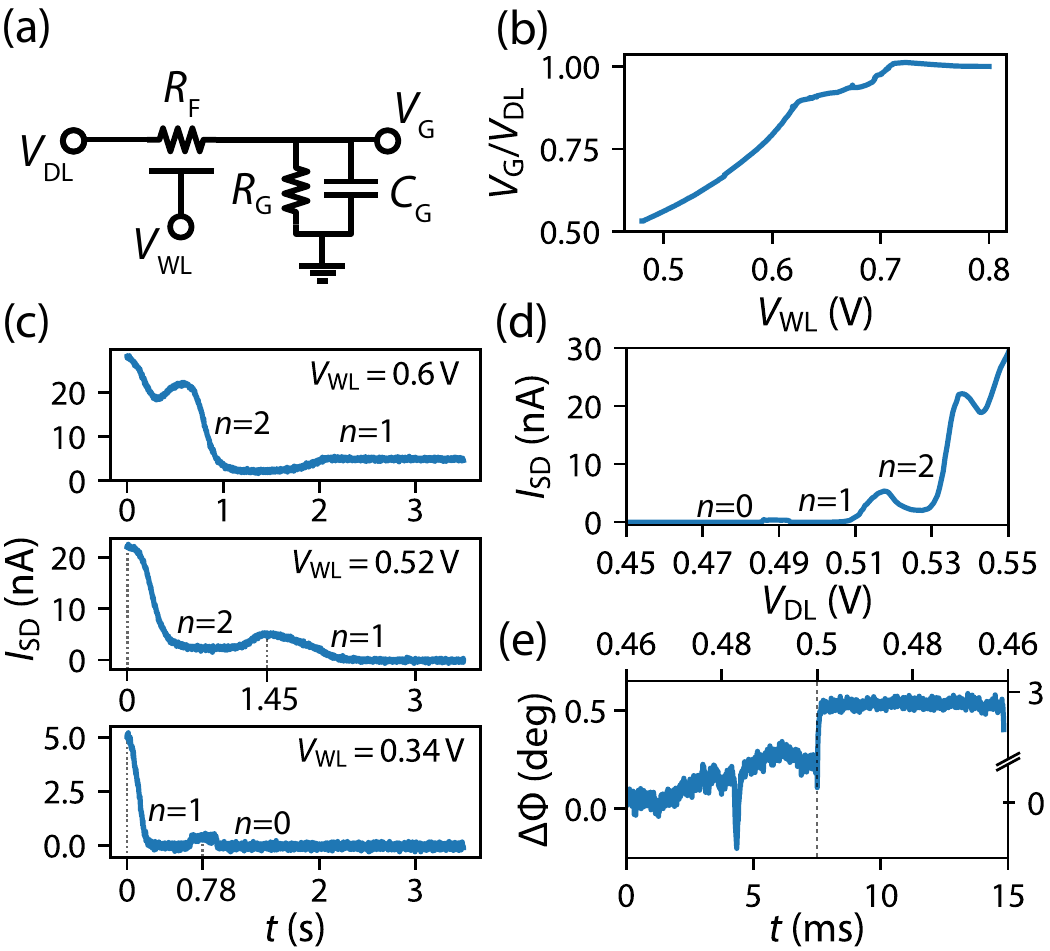}
  \caption{{\bf Charge retention time and fast switching}. \textbf{(a)} Equivalent circuit consisting of the variable control FET resistance $R_\mathrm{FET}$ and quantum device gate leakage $R_\mathrm{G}$ and capacitance $C_\mathrm{G}$. \textbf{(b)} Voltage divider characteristic of this circuit. \textbf{(c)} Demonstration of charge locking for different FET `OFF' states. Slow leakage of quantum dot gate charge is observed. \textbf{(d)} Quantum device transfer characteristic. \textbf{(e)} Demonstration of rf-sensing combined with fast switching of the control FET. Initially, the FET is biased above threshold and $V_\text{DL}$ is ramped from $0.46\, $V to $0.50\, $V. Tunneling of the first electron onto the quantum dot is observed (left axis). After $7.5\, $ms the FET is biased below threshold leading to a large jump in phase due to the change in resonance frequency (right axis) while $V_\text{DL}$ is ramped back down. In the `OFF' state no electron tunneling is observed.}
  \label{fig:switch}
\end{figure}

For multiplexing of the quantum device gate signal to be effective, the gate must be able to store the charge for a retention time which is long compared to the inverse of the refresh rate. %The retention time, a critical parameter that determines the number of qubits that can be subsequently read before a refresh voltage of the gate is required. 
To measure the charge retention time in our circuit, we study the dynamics of the quantum device when switching the control FET on and off. Measurements were performed in a different pair of devices, nominally identical to those used above. In Fig.\@ \ref{fig:switch}.(a) we present the equivalent circuit of the charge memory node, similar to a voltage divider for the dataline voltage $V_\mathrm{DL}$ with the variable channel resistance of the FET, $R_\mathrm{FET}$, and gate leakage resistance, $R_\mathrm{G}$, that represents dielectric losses through the gate-oxide. Both resistances determine the voltage $V_\mathrm{G} = \frac{R_\mathrm{G} }{ R_\mathrm{F} + R_\mathrm{G} } V_\mathrm{DL}$ appearing on the gate of the quantum device --- the capacitance of this gate, represented by $C_\mathrm{G}$, can be obtained from the gate voltage spacing $\Delta V_\text{DL}$ between consecutive Coulomb blockade oscillations plotted in Fig.\@ \ref{fig:switch}.(d). Using $C_\mathrm{G}^{n,n+1}=e/\Delta V_\mathrm{DL}^{n,n+1}$, where $n$ is the number of electrons in the dot, we obtain $C_\mathrm{G}^{0,1}=6.2\, $aF and $C_\mathrm{G}^{1,2}=7.0\, $aF.
%
% Fig.\@ \ref{fig:switch}.(d) which shows the turn on characteristic of the qubit with indicated Coulomb oscillations $C_\mathrm{G}^{n,n+1}=e/\Delta V_\mathrm{DL}^{n,n+1}$. We obtain $C_\mathrm{G}^{0,1}=6.2\, $aF and $C_\mathrm{G}^{1,2}=7.0\, $aF.
%
In Fig.\@ \ref{fig:switch}.(b) we show the voltage division $V_\text{G}/V_\text{DL}$ obtained by tracking the position of the Coulomb peak as a function of $(V_\text{WL}-V_\text{DL})$. We conclude that at $V_\mathrm{WL}< 0.5\, $V the resistance of the control FET channel becomes greater than the gate leakage in the quantum device.

The charging dynamics of the device is determined by the circuit $RC$ time constant $\tau=\frac{C_\mathrm{G} R_\mathrm{G} R_\mathrm{FET}}{R_\mathrm{G} + R_\mathrm{FET}}$. We study these dynamics by switching the control FET from an `ON' state to different `OFF' states and monitoring the resulting source-drain current through the quantum device (see Fig.~\ref{fig:switch}.(c)). In each case, $V_\mathrm{DL}$ is kept constant at 0.6~V.
%This is realistic as ideally different qubit devices fabricated using the same process should require similar tuning voltages.
By comparing the transient response with the quantum device transfer characteristic (Fig.~\ref{fig:switch}.(d)) we see that $I_\text{SD}(t)$ reproduces the Coulomb oscillations, with the steady-state current determined by the voltage divider and $V_\mathrm{DL}$.
For $V_\mathrm{WL}=0.6\, $V as the `OFF' state, $R_\mathrm{FET}< R_\mathrm{G}$ the discharging of the gate capacitor occurs mainly through the control FET channel. For a more resistive `OFF' state of the control FET, as given by $V_\mathrm{WL}=0.34\, $V, discharging of the gate capacitor occurs mainly through gate leakage since $R_\mathrm{F}> R_\mathrm{G}$ and the steady-state voltage on the quantum device gate $V_\mathrm{G}$ approaches zero.

Using the observed time dynamics of the current in Fig.~\ref{fig:switch}.(c), we characterize the single-electron retention time of the storage node through time lapses $\Delta t^{n,n+1}$ between successive Coulomb oscillations, obtaining $\Delta t^{1,2}=1450\, $ms and $\Delta t^{0,1}=780\, $ms.
These retention times can be used to estimate the following circuit parameters:
%and the resistance ratios in a time-dependent equation of the memory node, we estimate the circuit parameters. For the circuit resistances, we find $
$R_\text{F}(V_\text{WL}=0.52~V)=3.1\cdot 10^{18}\, \Omega$, $R_\text{F}(V_\text{WL}=0.34~V)=4.7\cdot 10^{18}\, \Omega$ and $R_\text{G}=3.5\cdot 10^{18}\, \Omega$.
For the RC time constant we find $\tau^{1,2},\tau^{0,1}\approx 12\, $s. 
These results provide valuable information to assess the suitability of time-multiplexing dispersive readout for large scale quantum computing. First of all, these values compare quite favorably to state-of-the-art DRAM cells, which show a leakage resistance on the order of $10^{15}\, \Omega$~\cite{DavidTaweiWang2005} and a refresh time of $64\, $ms~\cite{JEDEC2005}. Moreover, the retention times reported here are well above the typical expected readout times of $100\, $ns of gate-based reflectometry~\cite{Gonzalez-Zalba2015} and the single qubit coherence time of $28\, $ms in $^{28}$Si substrates~\cite{Veldhorst2015}. Considering typical operation times of spin qubits in silicon ($1\, \si{\micro s}$) this retention time will allow addressing of $10^6$ qubits before the voltage on one node needs to be refreshed. 

% Most importantly, from the middle and bottom panels we obtain the charge retention time between succesive we obtain a charge retention time which is on the order of $780$-$1450\, $ms until one electron will tunnel out of the quantum dot. This will determine the number of qubits which can be subsequently readout before a refresh of the gate voltages is required in order to maintain the electron occupation in the quantum dots. We can also relate this to a voltage drift of $16$-$40\, $mV/s.

% From the retention time and the voltage divider characteristic we estimate $R_\mathrm{F}$ and $R_\mathrm{G}$. We obtain a resistance larger than $R_\mathrm{G}=1\times 10^{18}\, \si{\ohm}$. (add exact values!).
% This immense resistance is valuable information to feed back to the transistor foundry as state-of-the-art DRAM cells show a leakage close to $1\, $fA which corresponds to a resistance of $10^{15}\, \si{\ohm}$ for a voltage difference of $1\, $V. (citation!)

As a demonstration of time-multiplexed dispersive readout, we perform a rf reflectometry measurement followed by fast switching of the control FET, shown in Fig.~\ref{fig:switch}.(d). In the first part of the measurement cycle, $V_\mathrm{DL}$ is ramped from $0.46\, $V to $0.50\, $V while the control FET gate is `ON' ($V_\mathrm{WL}=1.2$~V), leading to a tunneling of the first electron onto the quantum dot. Then, after $7.5\, $ms, the control FET is switched to the `OFF' state ($V_\mathrm{WL}=0.3$~V) and $V_\mathrm{DL}$ is ramped down to $0.46\, $V. As expected, no dispersive response from the quantum dot is measured during this time period, which could instead be used to measure another quantum device connected to the same dataline via a different control FET. %Due to the difference in resonance frequency between the on and off state of the FET there is a large offset in the phase when switching the FET into the off state. 
In this way, multiple qubits could be measured sequentially within the retention time of the charge storage circuit. 
% When note that when switching the FET back to the on state after this $15\, $ms sequence we expect to see an electron tunneling out of the quantum dot. Such a tunneling event was not observed in this experiment. This can be explained by a tunnel rate beyond our measurement bandwidth and sensitivity, demonstrating the current limits of time resolved gate-based charge sensing.

Although integration of quantum and classical CMOS devices promises major advantages in practical quantum computing architectures, for example in addressing wiring challenges, this comes at a cost of managing the dissipation of heat from the classical control circuits. We estimate the heat dissipation per device in our experiments, based on the dynamic power produced by the control FET which is given by $P=C_{\rm FET} f_{\rm op} {\Delta V}^2 $. We estimate $C_{\rm FET}$, the FET capacitance, to be $C_{\rm FET}=13\, $fF, given the FET dimensions ($50\, $nm $\times$ $10\, \si{\micro m}$ gate and $1.3\, $nm equivalent oxide thickness). The operation frequency $f_{\rm op}$ is limited by readout time, typically $t=1\, \si{\micro s}$ for rf-sensors, which determines the maximal frequency of $1$~MHz. The largest voltage difference between the `ON' and `OFF' state of the FET chosen in this experiment comes close to $\Delta V=1\, $V. From this we estimate a power dissipation of $P=13\, $nW per device, which can be treated as an upper bound as the dimensions and thus the capacitance of the FET, the operation frequency and voltage difference $\Delta V$ could all be reduced. Nevertheless, assuming a cooling power of $400\, \si{\micro W}$ at $100\, $mK, as achieved in current dilution refrigerators, operation of at least $30,000$ transistors would be possible at this temperature.

In conclusion, we have demonstrated the integration of three elements likely to play key roles in a large scale spin-based quantum computer: a quantum device (quantum dot), a classical control device (field-effect transistor) and sensitive charge readout (electrical resonator). Two of these have been fabricated on the same chip using CMOS technology and there is a potential for the $LC$ resonator to be made in a CMOS process. The footprint of such spiral inductors can exceed $500\times 500\, \si{\micro m^2}$~\cite{Burghartz1998}, however, by reducing the spiral dimensions~\cite{Hornibrook2013} or using kinetic inductance~\cite{Samkharadze2015} a two order of magnitude reduction in footprint area is expected. High quality factors could be achieved by using superconducting TiN, which is already found in the gate-stack of current CMOS transistors.
%showing the potential of this approach for large-scale integration. 
Overall, we have demonstrated a first step towards time-based multiplexing of gate-based radio-frequency reflectometry, with a charge sensitivity of $\delta q=165\, \si{\micro e  Hz^{-1/2}}$, motivating further experiments on multi-qubit circuits.
Another key area for further development will be incorporation of single-shot dispersive readout of single-electron transitions, as a necessary requirement for active feedback in quantum error-correcting protocols. 
%as well as the use of high-Q superconducting resonators\cite{Mason2010,Colless2013} and optimal rf-reflectometry circuit designs to improve the SNR to allow single-shot gate-based rf-reflectometry readout of quantum dots devices. JM COMMENT: I don't think this last part says very much, which is why I have commented it out.

\begin{acknowledgement}

We are grateful for useful discussions with A. Rossi and N. J. Lambert. This research has received funding from the European Union’s Horizon 2020 research and innovation programme under grant agreement No 688539 (http://mos-quito.eu) and Seventh Framework Programme (FP7/2007-2013) through Grant Agreement No. 318397 (http://www.tolop.eu.); as well as by the Engineering and Physical Sciences Research Council (EPSRC) through the Centre for Doctoral Training in Delivering Quantum Technologies (EP/L015242/1) and UNDEDD (EP/K025945/1). M.F.G.Z. acknowledges support from Hughes Hall, University of Cambridge. 
\end{acknowledgement}

% \begin{suppinfo}

% This will usually read something like: ``Experimental procedures and
% characterization data for all new compounds. The class will
% automatically add a sentence pointing to the information on-line:

% \end{suppinfo}

%\bibliography{Mendeley}

\providecommand{\latin}[1]{#1}
\makeatletter
\providecommand{\doi}
{\begingroup\let\do\@makeother\dospecials
	\catcode`\{=1 \catcode`\}=2\doi@aux}
\providecommand{\doi@aux}[1]{\endgroup\texttt{#1}}
\makeatother
\providecommand*\mcitethebibliography{\thebibliography}
\csname @ifundefined\endcsname{endmcitethebibliography}
{\let\endmcitethebibliography\endthebibliography}{}

\end{document}